# Current Lightweight Cryptography Protocols in Smart City IoT Networks: A Survey

M. Rana, Q. Mamun, R. Islam

*School of Computing and Mathematics, Charles Sturt University*
mrana@csu.edu.au, qmamun@csu.edu.au, mislam@csu.edu.au

**Abstract: With the advent of advanced technology, IoT introduces a vast number of devices connecting with each other and collecting a sheer volume of data. Thus, the demands of IoT security is paramount. Cryptography is being used to secure the networks for authentication, confidentiality, data integrity and access control. However, due to the resource constraint nature of IoT devices, the traditional cryptographic protocols may not be suited in all IoT environments. Researchers, as a result, have been proposing various lightweight cryptographic algorithms and protocols. In this paper, we discuss the state of the art lightweight cryptographic protocols for IoT networks and present a comparative analysis of the existing protocols. In doing so, this paper has classified the most current algorithm into two parts, such as symmetric and asymmetric lightweight cryptography. Additionally, we consider several recent developed block cipher and stream cipher algorithms. Furthermore, various research challenges of lightweight cryptography have been addressed.**

## 1. Introduction:

Internet of Things (IoT) is a cluster of physical nodes or objects or devices that communicate with each other across the Internet so the user could monitor, analyse, and control them remotely. In the last few years, IoT growth exponentially, and it occupies our lives in several areas such as city, agriculture, hospital, environment, homes, roads and so on. This expansion of the Internet of things making it the forthcoming future of the technology in human history. The IoT object usually equipped with different types of sensors and actuators, which collect numerous data and sent the accumulated data through the cyberspace for monitoring, analysing, controlling, and reaching various conclusions [1]. Most of these data are real-time data and help us to make the correct decision about the different service domains. However, this Internet-driven raw data need to be transferred securely and switched to human-understandable information so we could gather knowledge and use them in various domains such as smart city, agriculture, environment, interactive transport, and grid.

A Smart City is an efficient consumption of city resources cost-effectively for the urban ecosystem, which enhances the quality of people life [2]. The smart city provides facilities in a variety of service domain such as traffic, energy, environment, medicine, education, and safety. 70% of the services are currently provided on the three fields like traffic, safety, and power [3]. Figure 1 shows the leading smart city model around the globe. United Nations Population Fund presented more than half of the world's population now resides in the city. By 2050, it is expected to be increased to around 68% [4]. Cisco declared one billion dollar investment projects in smart cities in 2017. In China, more than 200 smart city developments are in progress [5]. However, smart city domains create numerous security and privacy challenges due to the weaknesses of each layer of a smart city system. Various attacks can reduce the quality of smart city facilities. In 2015, approximately 230 thousand Ukraine residents experienced a long period of power interruption as the electricity grid has been hacked by intruders hackers [6]. The enormous data that are shared in IoT-enabled environments can be exploited by malicious attacks that can be a security challenge of the smart city [7]. Hence, addressing and minimising of these security and privacy risks by the promotion of efficient security solutions is crucial for the success of smart city initiatives.

Ensuring privacy in smart city IoT nodes is challenging for several reasons. Firstly, the CPU in IoT devices is minimal and cannot compute complex algorithms [8-12]. Secondly, the power consumption of the security algorithm should be low since the majority of IoT devices work with a battery [9, 11-15]. Thirdly, simple sensors are connected to cover large physical network [14]. Finally, the cost of implementing the security algorithm should be little to deploy as many devices as possible [1, 10, 16, 17]. Conventional cybersecurity cryptography such as AES (Advanced Encryption Standard), RSA (Rivest-Shamir-Adleman), DES (Data Encryption Standard), blowfish, and RC6 cannot be used immediately to these smart domains because of the heterogeneity, scalability, and dynamic features of smart cities [15]. RC2 is the algorithm that consumes most power, on the contrary blowfish in the lowest one. Besides, most of these algorithms consume more energy while operating. Biswas has compared several WSN sensor motes and found that resource-constrained devices have as low as 2 kilobytes (kB) and 1 kB of Random Access Memory (RAM) and Electrically Erasable Programmable Read-Only Memory (EEPROM) correspondingly [15]. Such sensors cannot utilise the resource-consuming conventional security approaches [18]. Hence, secure communication is one of the significant concerns in low power and lossy systems which undoubtedly defines the necessity to develop Lightweight Cryptographic (LWC) algorithms for IoT security.

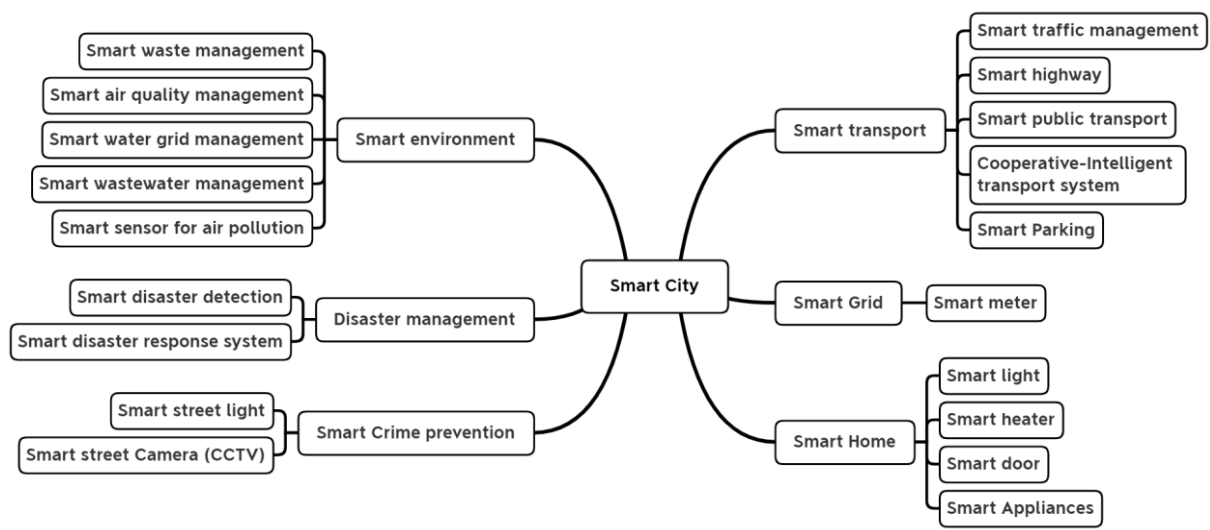

**Fig 1:** Major service in Smart City Network.

Despite the growing interest in IoT, the fundamental question is: What lightweight cryptography has developed to address the IoT security issues, which must be addressed to develop and execute to secure IoT networks efficiently? In an IoT environment, a systematic literature review on lightweight cryptography algorithm is essential to secure IoT communication. Hence, this paper focuses on the following main research questions:

1. What lightweight cryptography been has developed to address the IoT security issues?
2. How can a lightweight cryptography secure IoT structure?
3. What consequences the findings have on future IoT research?

This paper addresses the most current state of the art research in the field of lightweight cryptography covering the year 2019 and 2020. This paper also presents a comparative analysis of most present lightweight algorithms like LCC, LWHC, Modified PRESENT, SAT_Jo etc. Besides, the paper evaluates the most recent and existing protocols using a set of matrices such as block size, key length, gate area, technology value, round, latency, and throughput. The comprehensive evaluation demonstrates the requirements of lightweight cryptography ciphers. This paper organises into seven sections. Section I introduces IoT and the need for the development of LWC in the smart city. The IoT architecture and threats present in section II. Section III discusses IoT architecture and devices used according to the structure. Section IV describes security mechanisms in the IoT system. The most recent lightweight cryptography development in IoT is discussed in section V. Section VI presents the critical analysis of lightweight ciphers and the research gaps. Finally, the conclusion presents in section VII.

**2. IoT Architecture and threats**

This section draws the different layer of IoT architecture according to the devices functionality and various exposure to various attacks according to the different IoT layers. IoT domains show an enormous possibility. However, IoT network connects with heterogeneous devices with mixed OS and connect with a different communication protocol such as Wireless, Zigbee and mobile technology which creates considerable threats in security and privacy [19]. In this chapter, we draw an outline of IoT Architecture and discussion on different layers. Besides, we present various attacks in IoT according to the architectural layer.

IoT architecture contains four distinct critical layers: (i) Perception Layer, (ii) Network Layer, (iii) Middleware Layer, and (iv) Application Layer [9]. Table 1 demonstrates the IoT layer with the components of each layer and their tasks in details.

**Table 1:** IoT layer architecture and task accordingly.

| Layer | Component | Tasks |
|---|---|---|
| Application Layer | Third party application, Consoles, Websites, Touch panel. | Machine Learning, Business model, Graphs and Flow charts. |
| Middleware Layer | Vendor specific third application. | Machine Learning, Processing, Pre-processing, and real time action. |
| Network Layer | Nodes, Gateways, Firmware. | Transmit and process data, Device management, process, secure routing. |
| Perception Layer | Sensors (Temperature and Humidity), Actuators (Relays and motor) | Transfer data, Identity, Monitor, Acquisition and Action. |

*2.1 Application Layer and Security Attacks*

The application layer is the top layer in the IoT infrastructure. The Middleware layer passes information to this layer to process with different Applications. Application layer represents the IoT data as a business model, flowchart, and graph. Smart city, smart home and smart car are some of the examples of application layer automation. Some of the application layer attacks are buffer overflow attacks [20], Cross-site Scripting attack [20], SQL injection attack [21], denial of service attack [22], phishing attacks [23], and data privacy issue [24].

*2.2 Middleware Layer and Security Attacks*

This Middleware layer operates the vendor-specific services for various IoT node information which performs as a link between the Network and Application layer. This link facilitates to process, pre-process, and store IoT node information based on the third party and node requirement [20]. An intruder can introduce different attacks in the Middleware in a different way such as application security attack [20], unauthorised access attack [25], replay attack [21], sleep deprivation attack [22], data security attack [26] etc. Middleware and application layer both are using resource-rich devices which can use the traditional cryptography to secure the IoT networks. Consequently, these two layers devices are not focusing on this paper. Paper mainly discuss the security technique of resource-constrained devices of network and perception layers.

*2.3 Network Layer and Security Attacks*

Network layer also called Transit layer which processes and securely routes or transmits the data throughout the IoT infrastructure. This layer uses different protocols like Zigbee, Bluetooth, IR, and 6LowPan for data transmission. For further processing and action, Network Layer depends on the Middleware Layer. Following are the different attacks in this layer.

- Eavesdropping: Eavesdropping is a passive attack and extracts the message contents from network broadcastings. It snoops, captures and sniffs broadcasted data then initiate different attacks or still different critical information [27].

- Device Cloning attacks: As IoT devices are easily accessible from the network, an intruder can create a clone of the device and can compromise the IoT network infrastructure using these devices [27].

- Spoofing attacks: Things are connected to the network either directly or through a gateway in the IoT structure. An attacker can physically capture the node or gateways and can replace or reprogrammed with malicious code. Things and gateways must be authenticated, and the date should be encrypted to prevent this kind of attack [28].

- DDoS attacks: IoT devices are exposed as the IoT architecture usages heterogeneous, and resource-constrained nodes. Firstly, an attacker captures the device credentials and gets access to the gateways/devices. A hacker uses network information to explore the IoT devices and can initiate a DoS attack by sending fake packets and down the entire system [29].

- Key Attack: Some devices pre-shared keys are hard-coded within the code. Hence, the intruder can easily capture this information [30].

- Traffic analysis: Traffic analysis is another passive attack and able to retrieve valuable information from the network traffic such as source and destination details from the header of the transmitted communications [12].

- Brute-force attack: This known as an exhaustive search. This is a cryptographic hack which depends on predicting possible patterns of a targeted password to discover the correct password [31].

- Man-In-The-Middle attacks: Due to the heterogeneous current IoT architecture, an intruder captures communications between two parties to eavesdrop or modify traffic travelling between them secretly. Attackers may capture personal information, login credentials or disruption communications and corrupt data [32].

- Sinkhole attacks: An adversary generates sinkholes to attract traffic flow from all the IoT devices. The network traffic can later be rerouted to other devices than the destination gateway. This attack compromises the IoT devices privacy and confidentiality [33].

*2.4 Perception Layer and Security Attacks*

The crucial role of IoT system is to gather and transmit information from the real world. Henceforth, the perception layer possesses different kinds of data gathering, processing and transmitting devices like pressure sensors, temperature sensors, Bluetooth, ZigBee etc. The perception layer can be split into two parts, such as (a) perception node (sensors or controllers, etc.) and (b) perception network that communicates to the upper layer of IoT architecture [9]. Perception node like sensors and actuators collect and control information. However, perception network transmits the collected information to the gateway. Perception layer uses WSN, RFID and GPS technology.

In the perception layer, the node could be attacked or intruded or compromised physically. Generally, this compromised device called faulty nodes. In order to ensure the quality of service, it is necessary to detect the defective devices and take action to avoid further degradation of the service. A localised fault detection algorithm was delivered to discover the faulty nodes in WSN [9]. Da Silva et al. [34] suggested a decentralised intrusion discovery system paradigm for the wireless sensor network. Wang et al. [35] proposed the intrusion discovery probability in both homogeneous and heterogeneous WSN.

The cryptography cipher algorithms and key management scheme is used to secure perception layer network communication. Device authentication uses a private key algorithm which has greater scalability and can ensure the security of the system without complicated key management algorithm [36]. Key management involves a secret key generation, distribution, storage, updating, and destruction. Key distribution system divided into four categories such as a) key broadcast distribution [37], b) group key distribution [38], c) master key pre-distribution; [39] and d) pairwise key distribution [40, 41]. Perception layer vulnerable to the following attacks.

- Physical Damage: Generally, IoT devices are located in public places. Consequently, physical nodes could be captured, damaged, or compromised. An intruder can tamper the device and use this device to log into the IoT gateway, which they can modify and capture network traffic and other secret information [42].

- Code Injection attacks: As an intruder can get access to the physical IoT devices, they can manipulate the devices by introducing malicious codes into the devices [27].

- Jamming attacks: This is a widespread attack for perception layer devices. IoT edge devices use wireless protocol for network communication to transfer data to a different layer, program devices, receive instructions form the upper layer [41].

- Battery draining: An intruder objective to drain the IoT device battery. After capturing the device, an adversary continuously performs an energy-hungry operation. By doing this, an attacker can deteriorate the network by reducing the battery power of resource-constrained IoT sensors [43].

While there are various attacks as mentioned above, the researchers also suggested several mitigation techniques such as cryptography, authentication, securing physical devices etc. However, these attacks are still creating serious concern for low resource devices used in the Internet of things domains.

## 3. Devices in different IoT layers

IoT devices are present in all architectural layer which have a limited proficiency due to low memory, internal storage, computational capability and power. The IoT environment comprises of various service architectures, protocols and network design to deal with billions of IoT nodes to exchange information. IoT devices can be generally divided into three categories like Class 0, Class 1 and Class 2 [44, 45].

Class 0 or low-end IoT devices are with constrained resources like memory, power, and computational capability, which are mainly present at the first or perception layer of the IoT architecture. These low-end devices sense data and communicate with lightweight communication protocols. The RAM varies from 1 to 50 KB, and flash memory ranges from 10 to 50 KB [46]. Security is the primary concern in these low-end nodes as there are vulnerable to threats.

Class 1 or middle-end IoT devices have more resources compared to low-end nodes. These devices are basic microcontrollers and sit over low-end devices in IoT architecture to improve the abilities of class 0 nodes devices [47]. These devices have a higher clock rate like from 100 MHz to 1.5 GHz, and RAM varies from 100 KB to 100 MB. The flash memory varies from 10 KB to 100 MB. These devices can use data encryption technology to secure the data. Arduino, Netduino are some of the middle-end nodes and they also present at both first and second layer of IoT design [48, 49].

Single-board computers with a high number of resources in terms of CPU, RAM, flash memory is in Class 2. These devices support traditional operating systems such as LINUX, UNIX [50] and support growing technologies like artificial intelligence, machine learning, deep learning and neural network. These devices have comparatively less security concern due to the higher resources [51]. Table 2 demonstrates the comparison of communication network technology and interfaces of the different IoT devices [46].

**Table 2:** Comparison of IoT devices in terms of communication technology and interfaces.

| Device | Interface provided | Communication |
| --- | --- | --- |
| Raspberry Pi | HDMI, micro USB, USB 2.0, Ethernet, WLAN, Blueetooth4.2, CSI camera port, DSI display port | IEEE 802.11b/n/ac wireless LAN, Bluetooth 4.2, BLE and Gigabit Ethernet |
| Beagleboard | HDMI, 3.5 mm stereo in/out, I2X, UART, LCD | Ethernet, WLAN, Bluetooth |
| Netduino | UART, I2C, SPI | Ethernet, low power Wi-Fi |
| Arduino | USB, UART, ADC, I2C | Ethernet, IEEE 802.11ah, Wi-Fi support |
| OpenMote-CC2538 | I2C, SPI | IEEE 802.15.4 with 2.4 GHz band |
| TELOSB | I2C, SPI | IEEE802.15.4 at 250 Kbps rate |
| OpenMote-B | I2C, SPI, USB 2.0 | IEEE802.15.4 g |
| LSN50 | I2C, ADC, DAC, USART, USB | Wireless chip |

IoT devices have fundamental restrictions such as processing power, storage, memory, power consumption and connectivity [52, 53]. Lightweight cryptography needs to consider those limitations while designing and implementing an IoT network. The effectiveness of a network depends on design complexity, power consumption, throughput, and CMOS technology.

Design complexity is determined by the gate value (GE). Power intake is crucial for active devices like wireless sensor devices. Nevertheless, energy consumption is the principal interest of passive devices, such as RFID tags and smart cards. Energy use is directly related to a chip area [54]. A small area indicates low energy consumption. CMOS technology also affects performance qualities. Distinct technologies and standard-cell libraries generate different outcomes; for instance, the identical execution of PRESENT generates 1075GE on 0.18μm, 1169GE on 0.25μm and 1000GE on 0.35μm CMOS technology [55]. Table 3 presents the GE and power consumption in relation to CMOS technology [46]. Table 4 describes the frequency, energy, RAM and ROM features of different microcontrollers in the marketplace [56].

**Table 3:** Characteristics of different CMOS technology.

| CMOS technology node (μm) | Gate density (kGEs/mm$^2$) | Power consumption (nW/MHz/GE) |
|---|---|---|
| 0.35 | 6 | 18 |
| 0.18 | 125 | 15 |
| 0.13 | 206 | 10 |
| 0.09 | 404 | 7 |
| 0.065 | 800 | 6.68 |

**Table 4:** Features of various microcontroller platforms.

| Microcontroller platform | Frequency (MHz) | RAM (KB) | ROM (KB) | Power (mA) |
|---|---|---|---|---|
| 8-bit | 4-8 | 0.064-4 | 1.4-128 | 2.2-8 |
| 16-bit | 4-8 | 2-10 | 48-60 | 1.5-2 |
| 32-bit | 13-180 | 256-512 | 4000-32,000 | 31-100 |

## 4. Securing the IoT system

Section four concisely discusses lightweight algorithms use to secure the Internet of things network communication. Furthermore, this chaper classify the latest developed lightweight algorithm.Figure 2 illustrates a different type of most recent lightweight cryptography which primarily split into two categories symmetric and asymmetric algorithm. The symmetric lightweight algorithm further divided into Lightweight Block Cipher (LWBC) and Lightweight Stream Ciphers (LWSC). Elliptic curve cryptography (ECC) falls under symmetric cryptography. The factors of the lightweight cryptographic primitives are evaluated by the key size, block size, number of rounds, and structures. We will discuss the recent development of three cipher technology such as block cipher, stream cipher and elliptic curve cipher to secure resource constrained IoT network.

*4.1 Lightweight Block cipher*

Block cipher is one kind of symmetric ciphers, and a complete block of data is processed at the same time. Lightweight block ciphers are used in two distinct styles of networks like Substitution-permutation network (SPN) and Feistel network (FN). Feistel structure is a design of the same circuit for encryption and decryption with minimal overhead. For encryption and decryption procedure, Feistel structure uses the same program code, which ensures low memory requirement [57]. The SPN network is faster without a key schedule which makes the system vulnerable to attacks. SPN structure is more suitable because of lesser execution round requirement and has lower power expenditure [19].

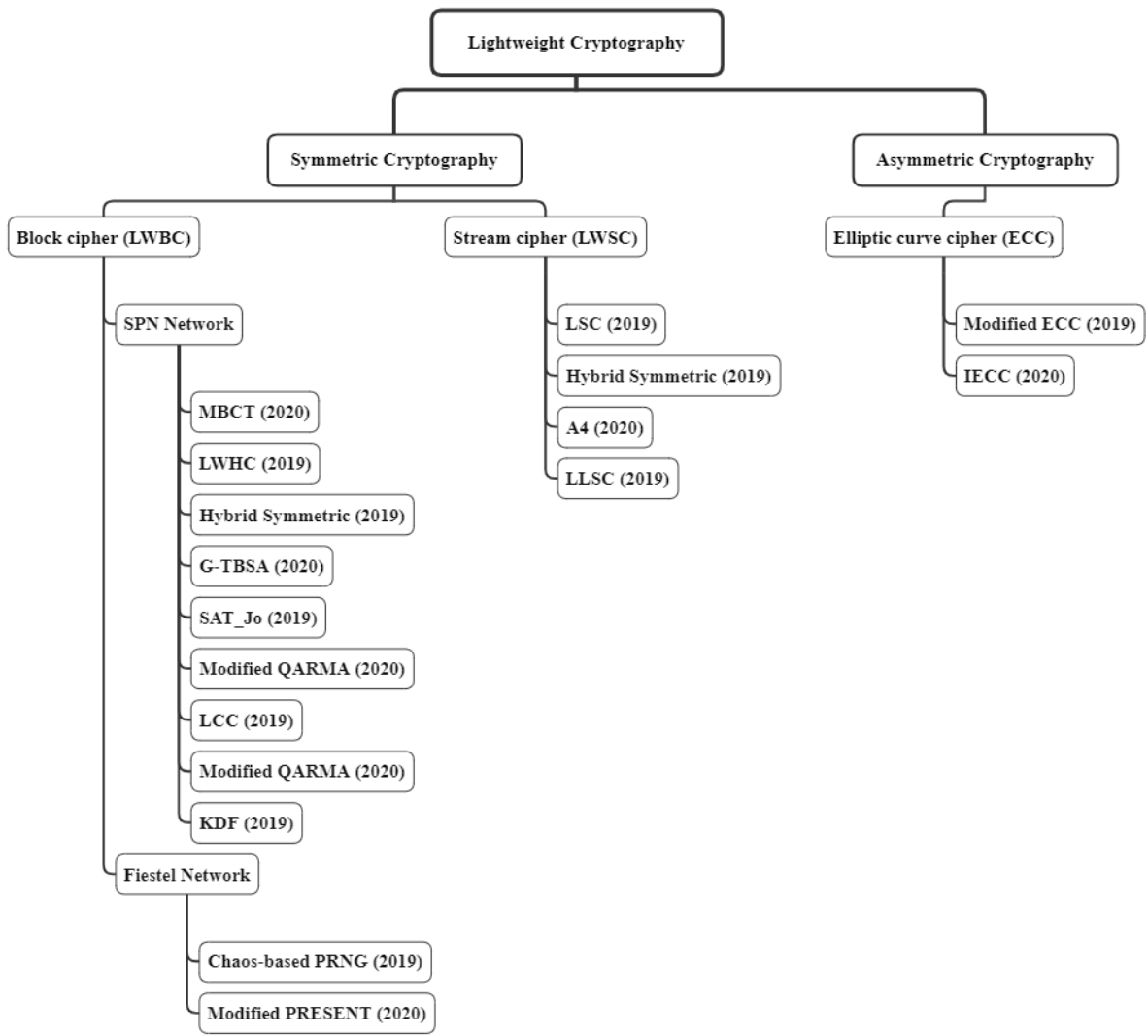

**Fig 2:** Classification of a recent developed lightweight cryptography algorithm.

Key size, block size, structure type, and the number of rounds is the primary considerations to evaluate a lightweight block cipher. In [58] writers recommend that a lightweight algorithm must focus on the three challenges such as a minimal memory, low power intake, sufficient security level. A Lightweight algorithm should have a small block size of 32–64 bits in comparison with the conventional 64 and 128 bits block size [59].

*4.2 Lightweight Stream cipher*

A stream cipher is another category of lightweight weight symmetric cryptographic algorithm and encrypts and decrypts data bit by bit. A stream cipher is simpler and quicker compared to block ciphers. This cipher is developed applying linear feedback shift registers (LFSRs) and nonlinear feedback shift registers (NLFSRs) [60]. It is extensively applied in WSN, cell phones etc. [60, 61]. Stream ciphers are used for both speed and fewer computation requirements. Some common stream ciphers are RC4, Salsa20, Trivium and Chacha.

*4.3 Lightweight Elliptic curve cipher*

Asymmetric ciphers like ECC also use to secure IoT network. ECC ensures authentication and confidentiality both [49]. Asymmetric ciphers use larger key size and more memory consumption, which make this cipher less popular in IoT security. RSA and ECC are two asymmetric cryptography can be used in IoT network security. ECC use lowered key size to show a similar security level compared to RSA. El-Gamal and Diffie-Hellman key algorithm can secure IoT system. AES [50] is 100–1000 times quicker than ECC on 8-bit microcontrollers according to execution time. ECC is the popular option to design to provide security in the IoT system.

**5. Recent lightweight cryptography for IoT security**

This section briefly reviews the most current lightweight cryptographic protocols to secure IoT network in a resource-restricted system.

Prakash, Singh, and Khatri [30] developed a new hybrid algorithm called Lightweight hybrid cryptography (LWHC) that used a combination of LED and PRESENT Cipher with a compact key scheduling algorithm SPECK. This system used RECTANGLE S-Box to make it faster and more robust. Encryption is done by the use of LED, PRESENT and RECTANGLE S-Box. However, the SPECK algorithm is also used for key scheduling. Proposed system used 64 bit of plain data to encrypt and perform an XOR operation with 128 bits schedule key. The LED cipher used the typical key algorithm which is not resistive to the key attacks. However, 128 bits SPECK key schedule algorithm is used in the proposed system, which gives it lightweight nature and secures from various key attacks. This advanced system used 64bit block plain text XOR with 128 bits key that maintains its robustness. The proposed algorithm is lightweight and secure to the various key attacks; however, not resistive to the other security attack.

Noura, Couturier, Pham, and Chehab [31] proposed Lightweight stream cipher scheme (LSC) which is based on the dynamic key-dependent method to reach high-level security. This system also includes a few simple operations to reduce the overhead, and cryptographic primitives change in a dynamic, lightweight manner for each input block. Security and performance study confirms that the proposed cipher attains a high level of effectiveness and robustness, which make it suitable for resource-restricted IoT devices. It uses 128, 192, and 256 bits secret key with a nonce. This algorithm is a combination of CR4, Pseudo-random number generator (PRNG), and Linear-feedback shift register (LFSR) with 12 to 14 rounds. This algorithm shows high periodicity, low energy consumption, low computational power and resistive to statistical, Algebraic, and brute-force attacks. However, it is not adequate for the disclosure and de-synchronisation attacks.

Shantha and L. Arockiam [60] designed SAT_Jo system which is based on the substitution-permutation network with a new lightweight block cipher which is appropriate for tag-based functions of the IoT. This system computes a 4 x 4 S-box by 24 order of the Galois field. This design is based on the SPN network of a block cipher with DES and PRESENT [62] involves in 31 rounds. Its usages 64 bits block and 80 bits key size. This system presents an adequate level of security to the resource-constrained nodes of tag-based applications. This proposal offers a better balance between performance, resource requirements and security for resource constrained IoT systems. These lightweight primitives do not support for an extensive range of functions and impose many challenges to the attacker. These lightweight algorithms are considered as weak.

Kubba and Hoomod [63] considered a Hybrid symmetric model based on the PRESENT (Block cipher) and Salsa20 (stream cipher) cryptography algorithms. This algorithm also used a 2D logistic map of a chaotic system to generate pseudo-random keys which produce more complexity. The main aim of this proposed algorithm is to enhance the complexity of the current PRESENT algorithm and keep the performance of computational operations as nominal as possible. The proposed algorithm uses a 64-bits symmetric block cipher algorithm and a 128-bits length key. Mainly the block PRESENT algorithm is considered for fast its performance [64]. Salsa20 has proposed because of its efficiency to use with constrained nodes [65]. Therefore, PRESENT and Salsa20 cipher algorithms meet the lightweight algorithms speed and complexity requirement. 20 rounds of PRESENT keys algorithms are used instead of the 31 rounds keys. However, Salsa20 cipher algorithm is used to generate keystream. The suggested algorithm has a significant level of randomness and demonstrates efficient performance with rapid execution times. Hence, the executed time of the proposed algorithm is a satisfied lightweight algorithm speed. However, the proposed algorithm introduces more complexity while preserving the computational speed at a minimum.

Noura et al. [66] proposed One round cipher (ORC) lightweight algorithm based on a dynamic structure with a single round roll. This model generates a dynamic key and then used to develop two robust substitution tables, a dynamic permutation table, and two pseudo-random matrices. This dynamic cipher structure uses a single round while providing a high level of randomness and security. Proposed chipper is resistant to statistical attacks which exhibit high randomness. OCR cipher shows a high level of sensitivity which protect it for key-related attacks.

The Generalised Triangle Based Security Algorithm (G-TBSA) [67] is designed by Ahmed et al., which is applied in wireless sensor networks (WSNs) with low power Wi-Fi. G-TBSA is a combination of resource-friendly data encryption and an efficient key generation mechanism. The key generation process is the heart of the algorithm since its usages fewer resources to generate the keys which minimise the complexity and provide energy efficiency. The proposed mechanism used the non-right-angle triangle-based method, and the output signal is used instead of time calculation. The proposed G-TBSA is more energy-efficient than other algorithms. However, this method works only for the sensors device.

Modified PRESENT [68] is a new lightweight PRESENT cipher presented Chatterjee and Chakraborty which has changed the original PRESENT cipher by reducing encryption round and modifying the Key Register. The key register is updated by encrypting its value by adding a delta value function of TEA (Tiny encryption algorithm), which is another lightweight cipher. The additional layer helps to reduce the PRESENT round from 31 to 25, which is the minimum round required for security. The efficiency of the proposed algorithm is increased by encrypting the key register. The proposed algorithm proves its superiority by analysing different software parameter analysis like N-gram, Non-Homogeneity, Frequency Distribution graph and Histogram. This algorithm shows better performance in terms of gate value. However, this algorithm has not tested for battery consumption.

Key-Dependent and Flexible (KDF) [69] proposal supports the logic and delivers a new lightweight cipher, with a simple round event and a dynamic key for every message. Subsequently, the proposed cipher utilised for real-time Multimedia applications with limited resources. This algorithm generates dynamic cryptographic primitives and performs the mixing of selected blocks in a dynamic pseudo-random manner. Accordingly, different plain-text messages are encrypted differently, and the avalanche effect is also preserved. This proposal shows the high level of immunity to the attacks such as statistical, differential and brute-force attacks. KDF need less computational complexity (less delay) than AES.

Roy, Rawat, and Karjee [12] proposed Lightweight CA (LCC), a lightweight cellular automaton (CA)-based cipher. This method encrypts information at the perception layer, which shows more efficient than some of the existing ciphers like DES, 3DES. This algorithm passes the randomness tests according to the National Institute of Standards and Technology (NIST). Besides, it passes all the DIEHARD tests which show the strong security feature of LCC. Modified QARMA [59] proposed by Zhao, Yan, and Li a part-iterative architecture for QARMA, which integrates encryption and decryption operations. Systems use on ASIC in CMOS 55 nm technology, where a maximum frequency of 666.67MHz is achieved. The results show that this model has achieved 54% area reduced and concurrently 25 times max frequency enhanced, and the throughput increases by 1.56times contrasted with the unrolled implementation. However, further modification needed to optimise the resource intake.

Chaudhary and K. Chatterjee [70] designed Modified Block Cipher Technique (MBCT) which is a combination of one Matrix Rotation, XoR and Expansion function. The encryption process primarily changes in Expansion and Round key creation function. The key length is 256 bits, and 256 bits block plain text used in this process. There is 32 rounds use in this algorithm. This algorithm needs less encryption and decryption time to compare to AES, DES and SIMON. Modified MBCT is also used less memory than AES, DES and SIMON.

Thangamani and Murugappan [44] proposes lightweight cryptography primitives by combining the differential logical pattern (DLP), S-box pattern generation and random key generation which decrease the time and memory complexity by decreasing the number of pattern structure elements. It uses 16 x 16 blocks of the input message. This LWC technique provides higher complexity to the scheme, which makes difficult for a hacker to capture the information. DLP uses less memory and time. S-box, the random key and input text or image generate the patterns. The DLP method is applied to encrypt the input blocks prior to transmit the message. The receiver generates the key for decryption by inversing the random key and S-box. Then, the DLP decryption technique is used to rebuild the initial information. The benefits of this work are reduced memory consumption and time complexity.

Hamzaab et al. developed Chaos-based PRNG [71] encryption cryptography to maintain patients' data confidentiality. This algorithm is used to keyframes immediately after extracting them using a video summarisation technique from video information. Symmetric block encryption system operates the proposed chaos-based algorithm with one set of confusion and diffusion processes. It uses a new Pseudo-Random Number Generator (PRNG) based on Zaslavsky chaotic and 2D logistic map. Probabilistic performance makes this system more appropriate for real-time. This method is effective to resist various attacks such as the differential, statistical, and exhaustive attacks to find the secret keys. This technique is fast and safer than another current algorithm.

Gyamf, Ansere, and Xu [72] proposed an algorithm by improving ECC which is containing a lightweight ECC based on the Diffie-Hellman key exchange technique and Advanced Encryption Standard (AES). Two sets public and private keys are generated from ECC Standard curves constructed on the key size of 256-bits and 512bits. Key K, generated by the cloud server, use to encryption and decryption applying AES with 10 different sets of data. Sensor information is encrypted at the IoT devices with a low key before it transmitted to the upper layer, which confirms primary security as internet service is used. The IoT-Edge obtains the generated Public Keys from the remote server, then extract and updates the different IoT devices. It ensures the highest security level as performs the higher standard of encryption and decryption. This proposed Modified ECC (MECC) has decreased the complexity of the conventional algorithms and considerably decreased the run time for heavy encryption. Due to this reason, this solution suitable for resource constraint IoT nodes.

Khan et al. proposed IECC [73] as a secure framework for authentication and encryption in IoT-based medical sensor information. The IECC is a curve based system that has a specific base point derived from functions of a prime number. This system mixes of biometric parameters and user credentials. An additional secret key is created to improve network security compared to typical ECC. A public key is generated to encrypt the information, and a private key is used to decrypt the data. The secret key is generated from the private key, public key and elliptic curve point. The system possesses security requirements like low encryption, decryption time and communication overhead. The mean encryption and decryption time of IECC is 1.032 and 1.004 µs correspondingly. This value is less than the ECC and RSA value. Statistical analysis also shows the strength of this scheme.

Mohandas et al. [74] introduced A4 stream cipher uses a Linear Feedback Shift Register (LFSR) and a Feedback with Carry Shift Register (FCSR). A4 shows higher security and easy to implement in different applications to secure data communication. A seed box containing 256 hexadecimal numbers each of 128 bits is established at both ends sender and receiver. After receiving the seed value, the LFER primarily clocks to set a number of times. This clocking of LFSR makes sure the second level security as an intruder is completely unaware of this calculation, and it secures the system. A4 is entirely protected to algebraic attacks due to the arrangement of LFSR and FCSR. This algorithm also shows resistive to brute force and differentials attacks.

New lightweight stream cipher (NLSC) [75] on Chaos stream cipher algorithm is a combination of a chaotic system and two Nonlinear Feedback Shift Registers (NFSRs) designed by Ding et al. 80 bits secret key has used in this algorithm. This algorithm is a combination of Logistic chaotic system, two 40-level NFSRs, and three multiplexers. The test results show that the stream cipher here has good cryptographic characteristics and resists statistical attacks. This algorithm is also suitable for resource-constrained devices.

## 6. Discussion and limitation of existing lightweight cryptography

Recently developed cryptography can split into two sections like symmetric and asymmetric. Symmetric ciphers use reduced key length compared to the asymmetric algorithm; thus, they are vulnerable to security because of less complexity nature. However, asymmetric ciphers use more complexity to secure the IoT network communication, but the larger key length makes them slower. Studying all these crucial considerations, it is a necessity to improve an algorithm which will use less power, decrease the complexity, take less time, deliver first and adequate security to the low-end IoT devices [76]. Block cipher and stream cipher represent to a symmetric algorithm whereas ECC represents the asymmetric cipher.

Table 5 shows a comparative analysis of some of the most recent (2019-2020) proposed protocols. These protocols have been briefly reviewed in this paper. LWHC, SAT_Jo, Modified PRESENT, LCC, Modified QARMA, DLP and MBCT are block cipher algorithm which is suitable for resource constrained device in an IoT environment. Modified PRESENT and LWHC both used 25 rounds of the algorithm; thus, they need less computational power. However, they are protective of key attacks but vulnerable to other attacks. On the other hand, MBCT applied 256 bits of key and 32 rounds of algorithm which require low memory but need to verify differential and linear cryptoanalysis attacks. Modified ECC and IECC are suitable for resource-constrained devices. However, IECC only applies for authentication purpose. One round ORC is a stream cipher algorithm which is resistive to statistical analysis, but latency is comparatively high. Hybrid symmetric with PRESENT and Salsa20 need less computational power. Nevertheless, more time is required to calculate (computational time) the algorithm. G-TBSA require low energy consumption but is suitable for wireless sensor networks only. LSC and KDF are suitable for stream cipher and use less computational and less power to generate the algorithm, but they showed less resistive to disclosure and de-synchronisation.

**Table 5.** A comparative analysis of the most recent Existing protocols

| Proposed Work | Algorithm, Tools and Techniques. | Key Size, Block size and round | Cipher and network Type | Features |
|---|---|---|---|---|
| LSC (2019) | CR4, PRNG, Dynamic key, XorShift64, LFSR, XOR. | Secret keys 128, 192, 256 nonce and dynamic key 512 bits, 12 to 14 rounds | Stream | High periodicity. Low energy and computational power needed. Resistive to statistical, Algebraic, and brute-force attacks. |
| Hybrid Symmetric (2019) | PRESENT and Salsa20, XOR, Chaotic system, Pseudo-random keys | 64 bits block, 128 bits key | Block | Work efficiently with fast executed time. The less computational power needed. |
| LWHC (2019) | LED, PRESENT RECTANGLE S-Box, XOR, SPECK key generation. | 64 bits block, 128 keys | Block | Robust to key attacks. |
| ORC (2019) | KSA, RC4, SHA-512. | One round | SPN, FN | Resistive to statistical analysis, visual degradation, sensitivity test. |
| SAT_Jo (2019) | PRESENT, DES, S-Box | 64 bits block, 80-bit key, 31 rounds | Block cipher, SPN | This algorithm offers a better balance between performance, resource requirements and security for resource constrained IoT systems. |
| G-TBSA (2020) | TBSA, Non-right-angle triangle. | | WSN | Low energy consumption. Suitable for wireless sensor networks. |
| Modified PRESENT (2020) | PRESENT, TEA, S-Box, P-Layer. | 64 bits plain text, 80 bits key, 25 rounds. | SPN, FN | This algorithm shows better performance in terms of gate value. |
| KDF (2019) | Dynamic key generation, SHA-512 | 64 bits Key, | | Effect of error propagation is limited to the byte. |
| LCC (2019) | CA, PRN (pseudo random number), RV512, GCA, non-linier. | | Block chipper | Resistive to brute-force, linear cryptanalysis, differential cryptanalysis attack. |
| Modified QARMA (2020) | QARMA, ASIC, CMOS 55 nm, S-Box, Boolean, Permutation, MixColumms. | 64 blocks, 27 rounds. | Block cipher, SPN | This algorithm reduced 54% of the area and simultaneously max frequency increases by 25x. |
| MBCT (2020) | Matrix location, XOR, Expansion function. | 256 bits Key. 256 block plain text. 32 rounds. | Block Cipher | Less encryption and decryption time required compared to AES, DES, and SIMON. |
| Chaos-based PRNG (2019) | PRING, Confusion and diffusion operation. 2-D chaotic system. | | Block Cipher, Symmetric | Resistive to various attacks like differential, statistical, and exhaustive attacks for finding secret keys. |
| MECC (2019) | ECC, AES, Diffie-Hellman | 256 bits and 512 bits | Asymmetric, ECC | It reduced the complexity of the conventional algorithms which make it suitable for resource constraint IoT nodes. |

| IECC (2020) | ECC, SHA-512, XOR, Secret key. | 512 bits | ECC | Statistical analysis shows the strength of this scheme. |
| A4 (2020) | LFSR, FCSR, XOR, Boolean. | 126 bits Key | Steam cipher | A4 is entirely protected to algebraic attacks and shows resistive to brute force and differentials attacks. |
| LLSC (2019) | NFSR | 80 bits key | Steam cipher | This algorithm resists to statistical attacks. |
| DLP (2019) | S-Box, Random Key | | Block Cipher | Reduce memory consumption. Reduce time complexity. |

Table 6 illustrates a comparison between various lightweight ciphers which influence the selection of ciphers for resource-restricted devices. Block size, key size, gate area, latency and throughput are the main parameter for lightweight primitives. Lightweight cryptography primitives performance is calculated by several matrices like ley size, rounds, latency, throughput and gate area etc.

**Table 6:** Comparison of the different lightweight algorithm in terms of key length, block size, technology value, GE, latency, throughput, and number of rounds.

| Algorithm | Key size (bits) | Block size (bits) | Gate Area (GE) | Technology value (μm) | No. of round | Latency (Cycle) | Throughput (Mbps) |
|---|---|---|---|---|---|---|---|
| AES | 128/192/256 | 128 | 2400 | 0.13 | 10/12/14 | 226 | 56.64 |
| PRESENT | 80, 128 | 64 | 2195 | 0.13 | 31 | 31 | 206 |
| HIGHT | 128 | 64 | 3048 | 0.13 | 32 | 34 | 188 |
| KTANTAN | 80 | 64 | 688 | 0.13 | 12/16/20 | 255 | 25.1 |
| LED | 128 | 64 | 1265 | 0.13 | 32 | 48 | 133.33 |
| RECTANGLE | 128 | 64 | 1787 | 0.13 | 25 | 26 | 246 |
| PRINCE | 128 | 64 | 3491 | 0.13 | 12 | 12 | 533 |
| SIMON | 96 | 48 | 763 | 0.13 | 32/52/72 | 304 | 15.8 |
| Piccolo | 80 | 64 | 1260 | 0.13 | 25 | 237.04 | 237 |
| SFN | 96 | 64 | 1876 | 0.18 | 32 | 1876.04 | 200 |
| SAT_Jo | 80 | 64 | 1167 | 0.13 | 31 | 1270 | 14.9 |
| Modified PRESENT | 80 | 64 | 1884 | 0.13 | 25 | | |
| QARMA | 64 | 64 | 17109 | 0.13 | 27 | 1 | 1705 |
| Modified QARMA | 64 | 64 | 7844 | 0.13 | 27 | 27 | 2667 |

**Key size:** Key length or key size indicates the number of bits in a key which used in cryptographic cipher techniques. The encryption strength depends on the complexity to discover the key. Encryption intensity is defined by the key size used in the encryption process. Longer keys deliver more robust cipher on the other hand; it requires more power and complexity requirement. AES, LED, RECTANGLE and PRINCE use 128 bit of key which make them unsuitable for resource-restricted devices. Nevertheless, Modified QARMA algorithm applies only 64 bits key size for encryption purpose. Besides, SAT_Jo, Modified PRESENT, Piccolo and KTANTAN are suitable for the IoT devices as they use 80 bits of key size. More power consumption test needs to be done for the modified PRESENT cipher.

**Block size:** Block cipher operates on a constant length of bits sequence. Block cipher cryptography uses the same size of input and output block. Bigger the block size requires more CPU and battery needed to secure the IoT network. Hence, a smaller block size algorithm is suitable for the IoT end devices. According to the compared algorithm table, AES uses the largest block size which is 128 bits (16 bytes) to secure the network and SIMON applies the lowest 46 bits of a block. However, most of the block cipher consume 64 bits (8 bytes) block range.

**Gate area and technology value:** Lightweight cryptography is measured by the Gate Equivalent (GE). Gate Equivalent indicates the physical area essential to execute the algorithm. A suitable lightweight primitive requires less gate area. According to the ISO/IEC standardised [77] lightweight cryptographic cryptography should have the GE value from 1000 to 2000. Table 6 indicates KATANTAN, LED, RECTANGLE, SIMON, Piccolo, SFN, SAT-Jo and Modified PRESENT are the lightweight algorithm. In the IoT environment, power consumption is a critical consideration. Energy can be assessed on Gate value (GE) and subsequent CMOS technology value [78]. When CMOS technology shifts form μm to nanometre (nm) gate intensity rises. Table 6 represents the corresponding technology value of 0.13 μm.

**Number of rounds:** A part of the key size, ciphers use round-based execution which makes the cipher extra secure. Larger key and more rounds transform the system safer [79] and use more energy and computational power. Cryptographic designing requires to decrease the necessary iterations round number. As a result, this will diminish the necessary resources and latency, which are both necessary to preserve the major functionality of IoT nodes [66]. The current typical ciphers are not fitted to these IoT nodes because a greater number of round repetitions is essential to achieve the preferred security [31]. AES, KTANTAN and PRINCE require less round, but gate area is higher than the lightweight algorithm. SFN and Sat_Jo are the lightweight algorithms though they use more round than AES. Piccolo is the suitable lightweight in terms of key length, round, gate area and throughput.

**Latency:** Latency is calculated as the time needed between the initial approach of encryption, and encrypted output is generated [80]. Latency represents the number of cycles. Latency is critical for real-time applications like smart city, smart transport, and smart grid etc. The IoT devices are resource restricted heterogeneous devices with high latency, low energy, minimal computation capability, and low throughput [12]; therefore, the minimum achievable latency is required. Low computational complexity is an essential prerequisite for an effective cipher system to ensure low latency which leads to low power and resources overhead [66]. Consequently, it is necessary to design new cryptographic algorithms with low latency and resources consumption. PRINCE, QARMA, and modified QARMA have the lowest latency, which indicates they are suitable for resource constrained devices. However, according to the author [58], further modification is needed for QARMA algorithm. AFN, SAT_Jo, AES and Piccolo are some of the block ciphers with higher latency that makes them vulnerable to several IoT attacks. KTANTAN and Piccolo show more latency which makes these algorithms inapplicable for the Internet of things devices.

**Throughput:** Throughput measures how much data is transferred from a source at a unit time. Throughput is higher in traditional cryptography; in contrast, several IoT applications expect moderate throughput [57]. SAT_Jo cipher facilitates high throughput and low latency. Besides, this algorithm needs less energy, and less area compares PRESENT block cipher, which makes it suitable for the lightweight block cipher [61]. However, LED, SIMON, SFN, and SAT-Jo use more algorithm, which makes them expensive in terms of computational power. Most of the algorithms show vulnerability to the various attacks.

The trade-off between performance, cost, and security in the resource constrained environment draw in figure 3. Kong [79] quoted a significant link between the cost, performance, and security of the IoT devices. The system speed is dynamically affected by the process platform. Compared to serial architecture, parallel arrangement enhances performance and decreases latency. Consequently, parallel structure and lower number of cryptographic rounds increase performance. Moreover, the system cost is directly associated with the selections of algorithms and performance. More cost is necessary for more robust and faster cipher [57].

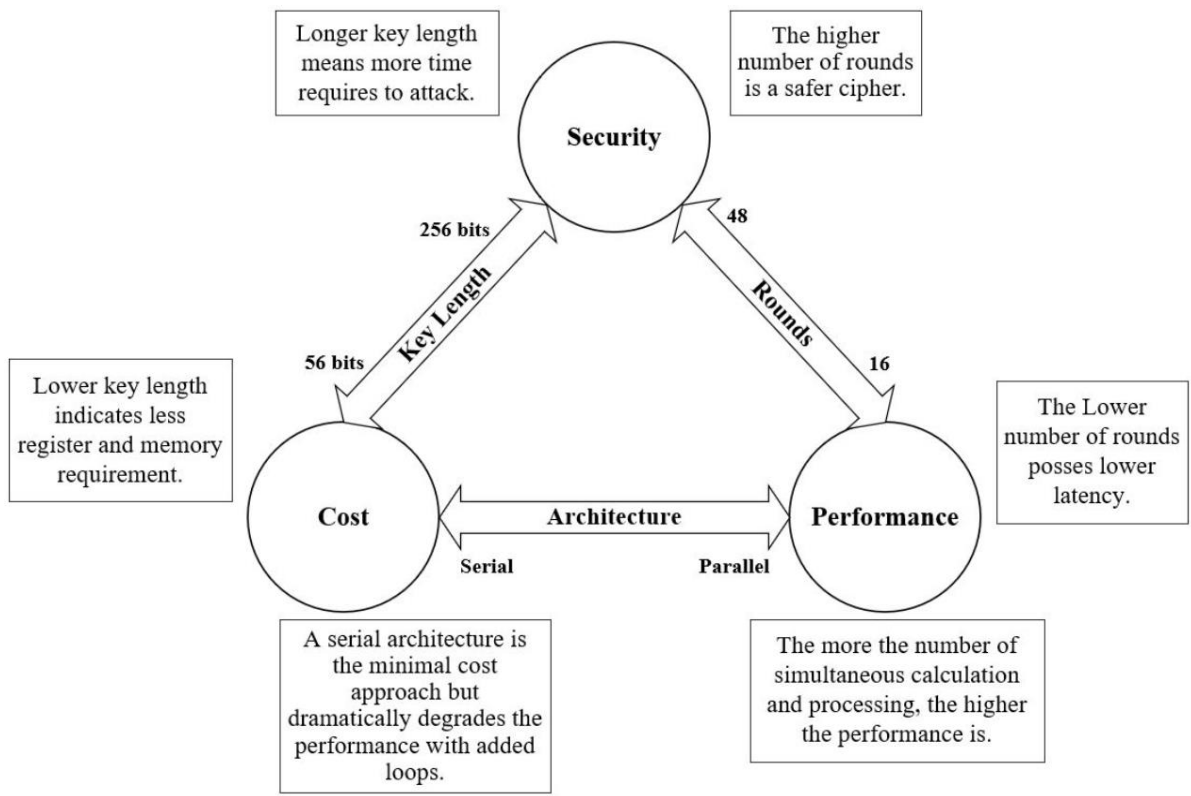

**Fig 3:** Trade-off between cost, performance, and security.

System performance is primarily computed by the number of rounds that a cipher can execute. More rounds involve more computational requirement; hence latency increases [79]. Conventional block cipher like AES uses a multi-round composition with several round repetitions which can be based on Substitution-Permutation Networks (SPN) or Feistel Networks (FN) [66]. Feistel network has a comparative advance than SP network considering the encryption and decryption activities. SPN consumes more resource than the Feistel network structure. However, the security of SPN is comparatively higher because the round function can modify all block messages in a reiterative round [81].

The key length is directly related to the security of lightweight cryptographic network communication. The larger the key more secure the network; however, larger keys require more memory and CPU. Consequently, makes the cipher unsuitable for resource-restricted devices. Table showing the largest key 256 bits and the lowest is 64 bits. QARMA uses the smallest key size, which makes them lightweight algorithm although their gate area is higher than the standard lightweight cipher. On the other hand, SIMON and KTANTAN gate are is in ultralow lightweight range and latency of KTANTAN is higher than the other.

Many new lightweight cipher algorithms have been proposed; nevertheless, the scope is there to improve in the area of security enhancement, decreasing latency, reducing energy consumption, lowering power consumption and chip area reduction. Different types of cipher are facing various challenges such as LCC shows resistance to various attacks, but key management strategies have not been developed yet. In contrast, G-TBSA consumes low energy, and this is suitable for wireless sensor network only. None of the modern lightweight algorithms is secure enough for both block cipher and stream cipher. Following are the various issues that need to address to develop lightweight block and stream cipher algorithm.

Challenges in lightweight block cipher:

- Implement a shorter key length.
- Generate simpler and fewer rounds of an algorithm.
- Use more frequent dynamic key.
- Apply smaller block of data.
- Develop a simple key structure

Issues in lightweight stream cipher:

- Apply smaller key size.
- Decrease the internal state.
- Reduce the chip area.

### 7. Conclusion

We have analysed recent research on lightweight cryptographic techniques used in IoT network to secure data communication. Each algorithm has some merits and demerits to ensure security while exchanging information in the IoT environment. Some algorithms demanded more storage space but fewer computations requirement and vice versa. Several algorithms are lightweight in terms of energy, computational power, and cost; nevertheless, they do not demonstrate resistance to different attacks. We found two types of algorithms according to the key arrangement such as symmetric and asymmetric cryptography. Popular recent symmetric algorithms used in Internet of things security are block cipher and stream cipher; however, none of them is ideal for securing resource constrained communication in IoT system. The security problem is a critical issue of IoT, which has not been appropriately addressed, and it is a crucial research topic in IoT network protection. A lightweight cryptographic algorithm needs to be developed to secure resource constrained IoT architecture. Growing attacks pattern to the IoT networks demands research on lightweight ciphers improvement. Future research can focus on reducing key size, using a more frequent dynamic key, decreasing block size, introducing more straightforward rounds, designing simple key schedules for lightweight block cipher development. Internal state, minimising key size, and initialising vector are some of the prime objectives to develop the lightweight stream ciphers.

### References

[1] A. Hameed and A. alomary, "Security Issues in IoT: A Survey," *2019 International Conference on Innovation and Intelligence for Informatics, Computing, and Technologies (3ICT),* 2019, doi: 10.1109/3ICT.2019.8910320. IEEE.

[2] D. Johnson and M. Ketel, "IoT: The Interconnection of Smart Cities," *SoutheastCon,* 2020, doi: 10.1109/SoutheastCon42311.2019.9020645. IEEE Xplore.

[3]  J. Lee, J. Kim, and J. Seo, "Cyber attack scenarios on smart city and their ripple effects," *International Conference on Platform Technology and Service (PlatCon),* 2019, doi: 10.1109/PlatCon.2019.8669431. IEEE.

[4]  "World Urbanization Prospects: The 2018 Revision," *Department of Economic and Social Affairs,* 2018, doi: United Nations, New York, USA. The United Nations.

[5]  Y. Li, Y. Lin, and S. Geertman, "The development of smart cities in China," *14th International Conference of Computer, Urban Planning and Urban Management,* pp. 7-10, 2015.

[6]  L. Cui, G. Xie, Y. Qu, L. Gao, and Y. Yang, "Security and Privacy in Smart Cities: Challenges and Opportunities," *IEEE Access,* vol. 6, pp. 46134-46145, 2018, doi: 10.1109/access.2018.2853985. IEEE Access.

[7]  J. Laufs, e. Borrion, and B. Bradford, "Security and the smart city: A systematic review," *Sustainable Cities and Society,* vol. 55, 2020, doi: 10.1016/j.scs.2020.102023. Elsevier Scince Direct.

[8]  X. Jiang, M. Lora, and S. Chattopadhyay, "An Experimental Analysis of Security Vulnerabilities in Industrial IoT Devices," *ACM Transactions on Internet Technology,* 2020, doi: doi.org/10.1145/3379542. ACM Digital Library.

[9]  Y. Yang, L. Wu, G. Yin, L. Li, and H. Zhao, "A Survey on Security and Privacy Issues in Internet-of-Things," *IEEE Internet of Things Journal,* vol. 4, no. 5, pp. 1250 - 1258, 2017, doi: 10.1109/JIOT.2017.2694844. IEEE.

[10] M. b. M. Noor and W. H. Hassan, "Current research on Internet of Things (IoT) security: A survey," *Computer Networks,* vol. 148, pp. 283-294, 2019, doi: 10.1016/j.comnet.2018.11.025. Elsevier.

[11] V. Rao and K. V. Prema, "Comparative Study of Lightweight Hashing Functions for Resource Constrained Devices of IoT " *4th International Conference on Computational Systems and Information Technology for Sustainable Solution (CSITSS),* 2019, doi: 10.1109/CSITSS47250.2019.9031038. IEEE.

[12] S. Roy, U. Rawat, and J. Karjee, "A Lightweight Cellular Automata Based Encryption Technique for IoT Applications," *IEEE Access,* vol. 7, pp. 39782 - 39793, 2019, doi: 10.1109/ACCESS.2019.2906326. IEEE Access.

[13] Q. Mamun, "A Qualitative Comparison of Different Logical Topologies for Wireless Sensor Networks," *Sensors,* 2012, doi: 10.3390/s121114887. sensors.

[14] A. Lepekhin, A. Borremans, I. Ilin, and S. Jantunen, "A systematic mapping study on internet of things challenges," *IEEE/ACM 1st International Workshop on Software Engineering Research & Practices for the Internet of Things (SERP4IoT),* 2019, doi: 10.1109/SERP4IoT.2019.00009. IEEE Digital Library.

[15] N. A. Gunathilake, W. J. Buchanan, and R. Asif, "Next Generation Lightweight Cryptography for Smart IoT Devices: Implementation, Challenges and Applications," *IEEE 5th World Forum on Internet of Things (WF-IoT),* 2019, doi: 10.1109/WF-IoT.2019.8767250. IEEE.

[16] K.-M. Chew, S. C.-W. Tan, G. C.-W. Loh, N. Bundan, and S.-P. Yiiong, "IoT Soil Moisture Monitoring and Irrigation System Development," *ICSCA 2020: Proceedings of the 2020 9th International Conference on Software and Computer Applications,* pp. 347-252, 2020, doi: 10.1145/3384544.3384595. ACM Digital Library.

[17] S. Zeadallya, A. K. Das, and N. Sklavos, "Cryptographic technologies and protocol standards for Internet of Things," *Internet of Things,* 2019, doi: 10.1016/j.iot.2019.100075. Elsevier.

[18] A. R. Sfar, E. Natalizio, Y. Challal, and Z. Chtourou, "A roadmap for security challenges in the Internet of Things," *Digital Communications and Networks,* vol. 4, no. 2, pp. 118-137, 2018, doi: 10.1016/j.dcan.2017.04.003. Science Direct.

[19] B. S. Sumit Singh Dhanda, Poonam Jindal, "Lightweight Cryptography: A Solution to Secure IoT," *Wireless Personal Communications,* 2020, doi: 10.1007/s11277-020-07134-3. Springer.

[20] V. Varadharajan, U. Tupakula, and K. Karmakar, "Study of Security Attacks against IoT Infrastructures," *Technical Report TR1: ISIF ASIA Funded Project,* 2018.

[21] M. Mahbub, "Progressive researches on IoT security: An exhaustive analysis from the perspective of protocols, vulnerabilities, and preemptive architectonics," *Journal of Network and Computer Applications,* vol. 168, 2020, doi: 10.1016/j.jnca.2020.102761. Elsiever.

[22] H. P. Alahari and S. B. Yelavarthi, "Performance Analysis of Denial of Service DoS and Distributed DoS Attack of Application and Network Layer of IoT," *Third International Conference on Inventive Systems and Control (ICISC),* 2019, doi: 10.1109/ICISC44355.2019.9036403. IEEE.

[23] S. N. Swamy, D. Jadhav, and N. Kulkarni, "Security threats in the application layer in IOT applications," *International Conference on I-SMAC (IoT in Social, Mobile, Analytics and Cloud) (I-SMAC),* 2017, doi: 10.1109/I-SMAC.2017.8058395. IEEE.

[24] A. Aggarwal, W. Asif, H. Azam, M. Markovic, M. Rajarajan, and P. Edwards, "User Privacy Risk Analysis For The Internet of Things," *Sixth International Conference on Internet of Things: Systems, Management and Security (IOTSMS),* 2019, doi: 10.1109/IOTSMS48152.2019.8939265. IEEE.

[25] B.-C. Chifor, I. Bica, and V.-V. Patriciu, "Mitigating DoS attacks in publish-subscribe IoT networks " *9th International Conference on Electronics, Computers and Artificial Intelligence (ECAI),* 2017, doi: 10.1109/ECAI.2017.8166463. IEEE.

[26] F. A. Bakhtiar, E. S. Pramukantoro, and H. Nihri, "A Lightweight IDS Based on J48 Algorithm for Detecting DoS Attacks on IoT Middleware," *IEEE 1st Global Conference on Life Sciences and Technologies (LifeTech),* 2019, doi: 10.1109/LifeTech.2019.8884057. IEEE.

[27] M. M. N. I. G.-M. J. Lloret, "Defenses Against Perception-Layer Attacks on IoT Smart Furniture for Impaired People," *IEEE Access,* vol. 8, pp. 119795 - 119805, 2020, doi: 10.1109/ACCESS.2020.3004814. IEEE.

[28] Y. M. Tukur and Y. S. Ali, "Demonstrating the Effect of Insider Attacks on Perception Layer of Internet of Things (IoT) Systems," *15th International Conference on Electronics, Computer and Computation (ICECCO),* 2019, doi: 10.1109/ICECCO48375.2019.9043248. IEEE.

[29] R. Kanagavelu and K. M. M. Aung. *A survey on SDN based security in internet of things, Advances in Intelligent Systems and Computing*, vol. 887, pp. 563-577, 2019.

[30] V. Prakash, A. V. Singh, and S. K. Khatri, "A New Model of Light Weight Hybrid Cryptography for Internet of Things," *2019 3rd International conference on Electronics, Communication and Aerospace Technology (ICECA),* 2019, doi: 10.1109/ICECA.2019.8821924. IEEE.

[31] H. Noura, R. Couturier, C. Pham, and A. Chehab, "Lightweight Stream Cipher Scheme for Resource-Constrained IoT Devices " *2019 International Conference on Wireless and Mobile Computing, Networking and Communications (WiMob),* 2019, doi: 10.1109/WiMOB.2019.8923144. IEEE.

[32] M. Gupta, M. Abdelsalam, S. Khorsandroo, and S. Mittal, "Security and Privacy in Smart Farming: Challenges and Opportunities," *IEEE Access,* Article vol. 8, pp. 34564-34584, 2020, Art no. 9003290, doi: 10.1109/ACCESS.2020.2975142.

[33] A. K. Mishra, A. K. Tripathy, D. Puthal, and L. T. Yang, "Analytical Model for Sybil Attack Phases in Internet of Things," *IEEE Internet of Things Journal,* vol. 6, no. 1, pp. 379 - 387, 2019, doi: 10.1109/JIOT.2018.2843769. IEEE.

[34] A. P. R. d. Silva, M. H. T. Martins, B. P. S. Rocha, A. A. F. Loureiro, L. B. Ruiz, and H. C. Wong, "Decentralized intrusion detection in wireless sensor networks," *1st ACM international workshop on Quality of service & security in wireless and mobile networks,* pp. 16-23, 2005, doi: 10.1145/1089761.1089765. ACM Digital Library.

[35] W. Yun, W. Xiaodong, X. Bin, W. Demin, and D. P. Agrawal, "Intrusion Detection in Homogeneous and Heterogeneous Wireless Sensor Networks," *IEEE Transactions on Mobile Computing,* vol. 7, no. 6, pp. 698-711, 2008, doi: 10.1109/tmc.2008.19. IEEE.

[36] Q. Jing, A. V. Vasilakos, J. Wan, J. Lu, and D. Qiu, "Security of the Internet of Things: Perspectives and challenges," *Wireless Network,* vol. 20, no. 8, pp. 2481-2501, 2014, doi: 10.1007/s11276-014-0761-7. Springer.

[37] S. C.-H. Huang and D.-Z. Du, "New constructions on broadcast encryption key pre-distribution schemes," *IEEE 24th Annual Joint Conference of the IEEE Computer and Communications Societies.,* vol. 1, 2005, doi: 10.1109/INFCOM.2005.1497919. IEEE Xplore.

[38] F. M.Al-Turjmana, A. E.Al-Fagihac, W. M.Alsalihb, and H. S.Hassanein, "A delay-tolerant framework for integrated RSNs in IoT," *Computer Communications,* vol. 36, no. 9, pp. 998-1010, 2013, doi: 10.1016/j.comcom.2012.07.001. Elsevier.

[39] M. Rana and Q. Mamun, "A robust and lightweight key management protocol for WSNs in distributed IoT applications," *International Journal of Systems and Software Security and Protection (IJSSSP),* vol. 9, no. 4, 2018, doi: 10.4018/IJSSSP.2018100101. IGI Global.

[40] H. Chan and A. Perrig, "PIKE: peer intermediaries for key establishment in sensor networks," *IEEE 24th Annual Joint Conference of the IEEE Computer and Communications Societies.,* 2005, doi: 10.1109/INFCOM.2005.1497920. IEEE Xplore.

[41] S. M. Tahsien, H. Karimipour, and P. Spachos, "Machine learning based solutions for security of Internet of Things (IoT): A survey," *Journal of Network and Computer Applications,* vol. 161, 2020, doi: 10.1016/j.jnca.2020.102630. Elsevier.

[42] K. Gafurov and T. M. Chung, "Comprehensive survey on internet of things, architecture, security aspects, applications, related technologies, economic perspective, and future directions," *Journal of Information Processing Systems,* Article vol. 15, no. 4, pp. 797-819, 2019, doi: 10.3745/JIPS.03.0125.

[43] V.-L. Nguyen, P.-C. Lin, and R.-H. Hwang, "Energy Depletion Attacks in Low Power Wireless Networks," *IEEE Access* vol. 7, pp. 51915 - 51932, 2019, doi: 10.1109/ACCESS.2019.2911424. IEEE Access.

[44] V. Vujović and M. Maksimović, "Raspberry Pi as a Sensor Web node for home automation," *Computers & Electrical Engineering,* vol. 44, pp. 153-171, 2015, doi: 10.1016/j.compeleceng.2015.01.019. ACM Digital Library.

[45] T. Kafer, S. R. Bader, L. Heling, R. Manke, and A. Harth, "Exposing Internet of Things Devices via REST and Linked Data Interfaces," *2nd Workshop Semantic Web Technologies for the Internet of Things,* 2017. Semantic Scholar.

[46] S. Bansal and D. Kumar, "IoT Ecosystem: A Survey on Devices, Gateways, Operating Systems, Middleware and Communication," *International Journal of Wireless Information Networks,* Article 2020, doi: 10.1007/s10776-020-00483-7. Springer

[47] S. Huh, S. Cho, and S. Kim, "Managing IoT devices using blockchain platform," *19th International Conference on Advanced Communication Technology (ICACT),* 2017, doi: 10.23919/ICACT.2017.7890132. IEEE.

[48] E. Baccelli *et al.*, "RIOT: an open-source operating system for low-end embedded devices in the IoT," *IEEE Internet of Things Journal* vol. 5, no. 6, pp. 4428 - 4440, 2018, doi: 10.1109/JIOT.2018.2815038. IEEE.

[49] D. Zhai, R. Zhang, L. Cai, B. Li, and Y. Jiang, "Energy-efficient user scheduling and power allocation for NOMA-based wireless networks with massive IoT devices," 2018.

[50] F. Shaikh, E. Bou-Harb, N. Neshenko, A. P. Wright, and N. Ghani, "Internet of Malicious Things: Correlating Active and Passive Measurements for Inferring and Characterizing Internet-Scale Unsolicited IoT Devices," *IEEE Communications Magazine,* vol. 56, no. 9, pp. 170-177, 2018, doi: 10.1109/MCOM.2018.1700685. IEEE.

[51] M. A. R. Shuman *et al.*, "Establishing groups of internet of things (IOT) devices and enabling communication among the groups of IOT devices," 2017.

[52] K. Fysarakis, G. Hatzivasilis, K. Rantos, A. Papanikolaou, and C. Manifavas, "Embedded Systems Security Challenges," *Measurable security for Embedded Computing and Communication Systems (MeSeCCS 2014),* 2014, doi: 10.5220/0004901602550266. Research Gate.

[53] C. Manifavas, G. Hatzivasilis, K. Fysarakis, and Y. Papaefstathiou, "A survey of lightweight stream ciphers for embedded systems," *Security and communicaiton netwoks,* vol. 9, pp. 1226-1246, 2015, doi: 10.1002/sec.1399. Wiley Online Library.

[54] A. Poschmann, "Lightweight Cryptography - Cryptographic Engineering for a Pervasive World," 2009. Ruhr-University Bochum.

[55] C. Rolfes, A. Poschmann, G. Leander, and C. Paar, "Ultra-Lightweight Implementations for Smart Devices – Security for 1000 Gate Equivalents," *International Conference on Smart Card Research and Advanced Applications,* pp. 89-103, 2008.

[56] R. Roman, C. Alcaraz, and J. Lopez, "A Survey of Cryptographic Primitives and Implementations for Hardware-Constrained Sensor Network Nodes," *Mobile Networks and Applications,* vol. 12, pp. 231-244, 2007, doi: 10.1007/s11036-007-0024-2. Springer Link.

[57] R. Kousalya and G. A. S. Kumar, "A Survey of Light-Weight Cryptographic Algorithm for Information Security and Hardware Efficiency in Resource Constrained Devices," *International Conference on Vision Towards Emerging Trends in Communication and Networking (ViTECoN),* 2019, doi: 10.1109/ViTECoN.2019.8899376. IEEE.

[58] N. Thangamani and M. Murugappan, "A Lightweight Cryptography Technique with Random Pattern Generation," *Wireless Personal Communications,* pp. 1409–1432, 2019, doi: 10.1007/s11277-018-6092-8. Springer.

[59] C. Zhao, Y. Yan, and W. Li, "An efficient ASIC Implementation of QARMA Lightweight Algorithm," *2019 IEEE 13th International Conference on ASIC (ASICON),* 2020, doi: 10.1109/ASICON47005.2019.8983618. IEEE.

[60] M. J. R. Shantha and L. Arockiam, "SAT_Jo: An enhanced Lightweight Block Cipher for the Internet of Things.," *2018 Second International Conference on Intelligent Computing and Control Systems (ICICCS),* 2019, doi: 10.1109/ICCONS.2018.8663068. IEEE.

[61] S. M. J. R, A. L, and S. K. Malarchelvi, "Security Analysis of SAT_Jo Lightweight Block Cipher for Data Security in Healthcare IoT," *ICCBDC 2019: Proceedings of the 2019 3rd International Conference on Cloud and Big Data Computing.,* pp. 111-116, 2019, doi: 10.1145/3358505.3358527.

[62] T. T. K. Hue, T. M. Hoang, and D. Tran, "Chaos-based S-box for Lightweight Block Cipher," *IEEE Fifth International Conference on Communications and Electronics (ICCE),* 2014, doi: 10.1109/CCE.2014.6916765. IEEE.

[63] Z. M. J. Kubba and H. K. Hoomod, "A Hybrid Modified Lightweight Algorithm Combined of Two Cryptography Algorithms PRESENT and Salsa20 Using Chaotic System," *2019 International Conference of Computer and Applied Sciences (1st CAS2019),* 2019, doi: 10.1109/CAS47993.2019.9075488. IEEE.

[64] W.-L. Cho, K.-B. Kim, and K.-W. Shin, "A Hardware Design of Ultra-Lightweight Block Cipher Algorithm PRESENT for IoT Applications," *Journal of the Korea Institute of Information and Communication Engineering,* vol. 20, no. 7, 2016, doi: 10.6109/jkiice.2016.20.7.1296.

[65] E. Lara, L. Aguilar, J. A. García, and M. A. Sanchez, "A Lightweight Cipher Based on Salsa20 for Resource-Constrained IoT Devices," 2018, doi: 10.3390/s18103326. Sensors.

[66] H. Noura, A. Chehab, L. Sleem, M. Noura, R. Couturier, and M. M. Mansour, "One round cipher algorithm for multimedia IoT devices," *Multimedia Tools and Applications,* 2018, doi: 10.1007/s11042-018-5660-y. Springer Link.

[67] S. F. Ahmed, M. R. Islam, T. D. Nath, B. J. Ferdosi, and A. S. M. T. Hasan, "G-TBSA: A Generalized Lightweight Security Algorithm for IoT," *2019 4th International Conference on Electrical Information and Communication Technology (EICT),* 2020, doi: 10.1109/EICT48899.2019.9068848. IEEE.

[68] R. Chatterjee and R. Chakraborty, "A Modified Lightweight PRESENT Cipher For IoT Security," *2020 International Conference on Computer Science, Engineering and Applications (ICCSEA),* 2020, doi: 10.1109/ICCSEA49143.2020.9132950. IEEE.

[69] H. Noura, A. Chehab, and R. Couturier, "Lightweight Dynamic Key-Dependent and Flexible Cipher Scheme for IoT Devices,," *2019 IEEE Wireless Communications and Networking Conference (WCNC),* 2019, doi: 10.1109/WCNC.2019.8885976. IEEE.

[70] R. R. K. Chaudhary and K. Chatterjee, "An Efficient Lightweight Cryptographic Technique For IoT based E-healthcare System," *2020 7th International Conference on Signal Processing and Integrated Networks (SPIN),* 2020, doi: 10.1109/SPIN48934.2020.9071421. IEEE.

[71] R. Hamzaab, Z. Yancd, K. Muhammade, P. Bellavistaf, and F. Titouna, "A privacy-preserving cryptosystem for IoT E-healthcare," *Information Sciences,* vol. 527, pp. 493-510, 2020, doi: 10.1016/j.ins.2019.01.070. Elsevier.

[72] E. Gyamf, J. A. Ansere, and L. Xu, "ECC Based Lightweight Cybersecurity Solution For IoT Networks Utilising Multi-Access Mobile Edge Computing," *Fourth International Conference on Fog and Mobile Edge Computing (FMEC),* 2019, doi: 10.1109/FMEC.2019.8795315. IEEE.

[73] M. A. Khan, M. T. Quasim, N. S. Alghamdi, and M. Y. Khan, "A Secure Framework for Authentication and Encryption Using Improved ECC for IoT-Based Medical Sensor Data," *IEEE Access,* vol. 8, pp. 52018 - 52027, 2020, doi: 10.1109/ACCESS.2020.2980739.

[74] N. A. Mohandas, A. Swathi, A. R., A. Nazar, and G. Sharath, "A4: A Lightweight Stream Cipher," *5th International Conference on Communication and Electronics Systems (ICCES),* 2020, doi: 10.1109/ICCES48766.2020.9138048. IEEE.

[75] L. Ding, C. Liu, Y. Zhang, and Q. Ding, "A New Lightweight Stream Cipher Based on Chaos," *Symmetry,* vol. 11, no. 7, 2019, doi: 10.3390/sym11070853. MDPI.

[76] S. Thapliyal, H. Gupta, and S. K. Khatri, "An Innovative Model for the Enhancement of IoT Device Using Lightweight Cryptography," *2019 Amity International Conference on Artificial Intelligence (AICAI),* 2019, doi: 10.1109/AICAI.2019.8701377. IEEE.

[77] C. Pei, Y. Xiao, W. Liang, and X. Han, "Trade-off of security and performance of lightweight block ciphers in Industrial Wireless Sensor Networks," *EURASIP Journal on Wireless Communications and Networking,* 2018, doi: 10.1186/s13638-018-1121-6. Springer Nature.

[78] G. Hatzivasilis, K. Fysarakis, I. Papaefstathiou, and C. Manifavas, "A review of lightweight block ciphers.," *Journal of Cryptographic Engineering,* pp. 141-184, 2018, doi: 10.1007/s13389-017-0160-y. Springer Link.

[79] J. H. Kong, L.-M. Ang, and K. P. Seng, "A comprehensive survey of modern symmetric cryptographic solutions for resource constrained environments," *Journal of Network and Computer Applications,* 2015.

[80] A. Hodjat and I. Verbauwhede, "Area-throughput trade-offs for fully pipelined 30 to 70 Gbits/s AES processors," *IEEE Transactions on Computers,* vol. 55, no. 4, pp. 366-372, 2006, doi: 10.1109/TC.2006.49. IEEE.

[81] L. Li, B. Liu, Y. Zhou, and Y. Zou, "SFN: A new lightweight block cipher," *Microprocessors and Microsystems,* vol. 60, pp. 138-150, 2018, doi: 10.1016/j.micpro.2018.04.009.